\documentstyle[twoside,epsfj,titlepage,12pt]{article}
\def\^{\wedge}

\def\paren#1{\left( #1 \right)}

\def\bra#1{\left[ #1 \right]}

\newcommand{\gtsima}{$\; \buildrel > \over \sim \;$}
\newcommand{\ltsima}{$\; \buildrel < \over \sim \;$}
\newcommand{\simgt}{\lower.5ex\hbox{\gtsima}}
\newcommand{\simlt}{\lower.5ex\hbox{\ltsima}}

\newcommand{\himpc}{{\hbox {$h^{-1}$}{\rm Mpc}} }

\newcommand{\GRB}{ {\scriptscriptstyle {GRB}} }

\font\smbf=cmb10 scaled\magstep1
\font\smit=cmti12
\begin{document}
\def\pp{\par\parshape 2 0truecm 16.5truecm 1truecm 15.5truecm\noindent}
\renewcommand{\thesection}{\normalsize\bf\arabic{section}.}
\renewcommand{\thesubsection}
{\normalsize\it\arabic{section}.\normalsize\it\arabic{subsection}}
\makeatletter
\def\@cite#1#2{\leavevmode\hbox{$^{\mbox{\the\scriptfont0 #1}}$}}
\makeatother
\begin{flushright}
TIT/HEP-268/COSMO-47 \\
UTAP/189-94 \\
September \ 1994 \hspace{1.7cm} \ \\
\end{flushright}

\begin{center}
{\Large {\bf Angular two-point correlation functions \\
for cosmological gamma-ray burst model \\ }}

\vspace{2cm}

{ {\sc Shiho Kobayashi, Shin Sasaki}}

{\it  Department of Physics, Tokyo Institute of Technology,
Oh-okayama, Tokyo 152 , Japan.}

and 

{ {\sc Yasushi Suto}}

{\it  Department of Physics, The University of Tokyo, 
Bunkyo-ku, Tokyo 113, Japan}

\vspace{1.5cm}

Submitted to Int. J. Mod. Phys. D

\vspace{1.5cm}

{\bf Abstract}

\end{center}

We compute the angular two-point correlation functions of the
gamma-ray bursts at cosmological distances.  Since the gamma-ray burst
emission mechanism is not yet established, we simply assume that the
gamma-ray burst sources are associated with high-redshift galaxies in
some way. Then on the basis of several simple models for the evolution
of galaxy spatial correlations, we calculate the amplitude of angular
two-point correlation functions on scales appropriate for the Compton
Gamma Ray Observatory data. we find that in most cases the predicted
correlations are difficult to detect with the current data rate and
the angular resolution, but models in which the bursts preferentially
occur at relatively low redshift ($z \simlt 0.5$) predict correlation
amplitudes on $\theta \sim 5^\circ$ which will be marginally
detectable with the Gamma Ray Observatory data in several years.  If
future observations detect a signal of angular correlations, it will
imply useful information on the correlation of galaxies at high
redshifts provided that the gamma-ray bursts are cosmological.

\vspace{0.5cm}
\noindent
PACS number(s): 95.85.Pw, 98.62.Py, 98.80.Es

\newpage

\section{\smbf Introduction}

The origin of the gamma-ray bursts (GRBs) remains one of the
challenging puzzles in high-energy astrophysics.  Although several
models to account for the violent energy production have been proposed
so far, none of them seems to be successful in explaining various
observed features of the GRBs including typical light curves, energy
spectra, source distributions, in a consistent and convincing
manner\cite{bla}.  Nevertheless the high degree of isotropy in the
angular distribution of GRBs observed by the BATSE (Burst and
Transient Source Experiment) on the {\it Gamma Ray Observatory} (GRO)
can be interpreted as a strong case for the cosmological origin
whatever the emission mechanism is\cite{pac}.  In particular, a
$V/V_{\rm max}$ test is a very stringent argument in favor of
cosmological origin\cite{mao}.

 If the GRBs are cosmological, the most natural idea is that the GRB
sources are associated with galaxies at redshifts around and up to $z
= (1\sim2)$\cite{pac,mao}.  If this is the case, the angular
distribution of the GRBs should trace that of distant galaxies,
provided that the GRB event rate be constant in time and insensitive
to the host galaxies.  Therefore the cosmological model for the GRBs
implies that the accumulating number of the GRBs detected by GRO
should eventually reveal the underlying angular correlation which is
expected for the distant galaxies.

In the present paper, we employ several simple models for the
evolution of galaxy correlation functions, and compute the angular
two-point correlation functions of the GRBs assuming that the GRB
event rate is constant.  We find that some models predict correlation
amplitudes on $\theta \sim 5^\circ$ which will be marginally
detectable with the GRO data in several years.  The predicted
amplitude would become significantly larger if we assume that the GRBs
are preferentially associated with distant rich clusters as
preliminarily reported by Cohen, Kolatte and Piran\cite{coh}.

The rest of the paper is organized as follows. In \S 2 our model for
the evolution of galaxy correlations and the GRBs is presented.
\S 3 describes the predictions of the angular two-point
correlation $w_{GRB} (\theta)$ for the GRBs, and examines the
dependence on the several model parameters. Finally \S 4 discusses the
detectability of the correlation, and the implications of the present
results.

\section{\smbf Models}

Throughout the present paper, we adopt a simple hypothesis that GRBs
are of cosmological origin. To be more specific, we assume that they
are 
associated with distant galaxies and that their event rate is
independent of host galaxies, for simplicity.  With these assumptions,
our present model is basically specified by two quantities; one is the
probability of the GRB event per galaxy per unit time,
$\phi(z)$, and the other is the spatial two-point correlation function
of galaxies, $\xi_{gg} (x,z)$.

Since no specific model for the GRB sources is 
available which
enables to predict the form of $\phi(z)$, we consider two simple
cases. Model A assumes that the GRB event rate is constant from some
critical redshift $z_c$ until the present:
\begin{eqnarray}
\phi^{A}(z(t))&=&\cases{
       \phi^{A}_0 & ; $0 \le z \le z_{c}$ \cr
         0        & ; $z_{c} \le z      $ \cr
                     }.             \label{step}
\end{eqnarray}
On the other hand, our model B assumes that the GRBs occur for a
finite period $\Delta z$ centered on $z_c$:
\begin{eqnarray}
\phi^{B}(z(t))&=&\cases{
       \phi^{B}_0 & ; $z_c-\Delta z/2 \le z \le z_{c}+\Delta z/2$ \cr
         0        & ; otherwise      \cr
                     },              \label{delta}
\end{eqnarray}
where $ {\phi_0}^A $ and $ {\phi_0}^B$ are constant and independent of
the redshift and host galaxies.

The other important ingredient in computing the angular correlation of
the GRBs is $\xi_{gg} (x,z)$. Once dark matter and cosmological
parameters are specified, it is fairly straightforward to
theoretically predict the two-point correlation functions of {\it dark
matter}. The two-point correlation function of {\it galaxies},
however, is quite difficult to estimate due to various complicated
physical processes involved in galaxy formation.  Therefore, we adopt a
power-law fit to the present galaxy-galaxy correlation
function\cite{dav}:
\begin{equation}
\label{poxi}
\xi_{gg}(x,z=0) = (x/x_0)^{-\gamma},
\quad x_0=5.4\himpc, \quad \gamma=1.77 ,
\end{equation}
where $h$ is the Hubble constant $H_0$ in units of $100$ km/sec/Mpc.
As for the time evolution, we assume that $\xi_{gg}
\paren{x,z}$ increases the amplitude independently of $x$:
\begin{equation}
\xi_{gg} \paren{x,z}=g^2(z,\Omega_0,\lambda_0) \xi_{gg}(x,z=0)
\label{power}
\end{equation}
where $\Omega_0$ and $\lambda_0$ denote the cosmological density
parameter and a dimensionless cosmological constant at present, and
$g(z,\Omega_0,\lambda_0)$ is a growth rate of the correlation
function.  As a growth rate $g(z,\Omega_0,\lambda_0)$, we use either
the linear growth rate
$D_1(z,\Omega_0,\lambda_0)/D_1(0,\Omega_0,\lambda_0)$, or the
power-law parameterization $(1+z)^{-(3+\epsilon-\gamma)/2}$.  The
parameter $\epsilon$ 
is introduced in order to incorporate 
the possible nonlinear growth in
a very crude manner.  If $\epsilon = 0$, the clustering measured in
proper coordinates is not changing, and if $\epsilon
=-3+\gamma=-1.23$, that measured in comoving coordinates is not
changing. 
Linear growth rate in the Einstein -- de Sitter models
corresponds to $\epsilon = \gamma - 1$

\section{\smbf Angular two-point correlation function of GRBs}

\subsection{\smit basic equations}

We consider the Robertson-Walker metric for the background universe:
\begin{equation}
ds^2=-dt^2+a^2(t)\bra{\frac{dx^2}{1-Kx^2}+x^2d\theta^2+
x^2\sin^2\theta d\phi^2},
\label{metric}\\
\end{equation}
where $a(t)$ is the cosmic scale factor, $K$ is the spatial curvature
constant which is related to the cosmological parameters as follows:
\begin{equation}
K=\frac{a_0^2 H_0^2}{c^2}(\Omega_0+\lambda_0-1),
\label{curvature}
\end{equation}
where $a_0 \equiv a(t_0)$ is the scale factor at the present time
$t_0$, and $c$ is the speed of light.  The universe is assumed to be
dominated by non-relativistic matter.  In order to examine the effect
of the background universe model, we compute the angular correlation
functions of the GRBs, $w_{GRB} (\theta)$, both in open universes
without cosmological constant ($\Omega_0<0$, $\lambda_0=0$) and in
spatially flat universes ($\Omega_0<0$, $\lambda_0= 1-\Omega_0$).

Using the small separation approximation (e.g., \S 56 in
Ref. 
5
) with the above assumptions, $w_{GRB} (\theta)$ is
related to $\xi_{gg}$ as
\begin{equation}
w_\GRB(\theta)= {\displaystyle 
{\int^\infty_0 x^4 dx {\phi^2(z) \over F^2(x)} \,
     \int^\infty_{-\infty} du \xi_{gg}(x,z) } 
\over
{\displaystyle 
 \left[\int^{\infty}_{0}dx x^2 {\phi(z) \over F(x)} \right]^2 
}},
\label{wxi} 
\end{equation}
where $u \equiv F(x) \sqrt{r^2/a^2 - x^2 \theta^2}$, and $F(x) \equiv
\sqrt{1-Kx^2}$.  In the above expression, the redshift $z$ should be
regarded as a function of the comoving coordinate $x$ through the null
geodesic equation. More explicitly, they are related to each other as
\begin{equation}
y \equiv {H_0a_0 x \over c} =
\frac{2[(\Omega_0-2)(1+\Omega_0z)^{1/2}+2-\Omega_0+\Omega_0z]}
{\Omega_0^2(1+z)} \label{yz2}
\end{equation}
for $\lambda_0=0$, and
\begin{equation}
y =\int^z_0 
\frac{dz^{\prime}}{\sqrt{\Omega_0(z^{\prime}+1)^3+1-\Omega_0}}
\label{yz3}
\end{equation}
for $\lambda_0 = 1-\Omega_0$.

For the power-law model for $\xi_{gg}(x,z)$ (eq.[\ref{power}]),
the integration over $u$ in eq.(\ref{wxi}) can be performed
and $w_\GRB(\theta)$ in this model reduces to\cite{lss}
\begin{eqnarray}
w_\GRB(\theta) &=& A \ \theta^{1-\gamma}, \label{powerwxi}\\
A &\equiv& H_\gamma \left({r_0 \over a_0}\right)^\gamma
\frac{\displaystyle \int^\infty_0 dx \, {x^{5-\gamma} \over F(x)}
     \phi^2(z) (1+z)^{\gamma} g^2(z,\Omega_0,\lambda_0)}
{\displaystyle \left[\int^{\infty}_{0}dx {x^2 \phi(z) \over F(x)}\right]^2},
\label{la}
\end{eqnarray}
where $H_{\gamma}$ is a product of Gamma functions; $H_{\gamma}\equiv
\Gamma(1/2)\Gamma\paren{(\gamma-1)/2}/\Gamma(\gamma/2)$.  The
cosmological model enters through the volume factor $F$, the relation
between $x$ and $z$ (eqs.[\ref{yz2}] and [\ref{yz3}]), and $ g(z,
\Omega_0, \lambda_0) $.

\subsection{\smit the Einstein -- de Sitter model}

In the Einstein -- de Sitter model with the power-law
$\xi_{gg}(x,z)$, the amplitude $A$ of $w_\GRB(\theta)$ (eq.[\ref{la}])
can be expressed in terms of the hypergeometric function ${}_2F_1$:
\begin{eqnarray}
A&=&
H_{\gamma} ({H_0r_0 \over c})^\gamma {9 \over 6-\gamma} \times
\cases{
   y_c^{-\gamma} \tilde f(y_c/2)   & ; (model A) \cr
{\displaystyle {y_+^{6-\gamma} \tilde f(y_+/2) - y_-^{6-\gamma} \tilde f(y_-/2)
   \over (y_+^3-y_-^3)^2} } & ; (model B) \cr
       } , \\
\tilde f(y) &\equiv& 
 {}_2 F_1(6-\gamma,2\gamma-2\epsilon-6,7-\gamma;y) , \\
y_c &\equiv& 2[1-(1+z_c)^{-1/2}] ,\\
y_\pm &\equiv& 2[1-(1+z_c \pm \Delta z/2)^{-1/2}] .
\end{eqnarray}
When $\Delta z \ll z_c \ll 1$, the above expressions reduce to
\begin{eqnarray}
A&\simeq&
H_{\gamma} ({H_0r_0 \over c})^\gamma \times
\cases{
{\displaystyle  {9 \over 6-\gamma} z_c^{-\gamma}} & ; (model A) \cr
{\displaystyle  {z_c^{1-\gamma} \over \Delta z}   } & ; (model B) \cr
       } .
\end{eqnarray}

Figure 1 plots the amplitude of $w_\GRB$ at $\theta =5^\circ$ as a
function of $z_c$ in the Einstein --  Sitter model.  
The angular scale $\theta =5^\circ$ roughly corresponds to the angular
resolution for the GRO. Thus mainly we refer to the amplitude at this
scale in the analysis below.
The angular correlation $w(5^{\circ})$ decreases with $z_c$ since the
number of accidental neighbors from the foreground and background
increases with $z_c$ and thus the angular correlation is reduced.  The
difference of evolution of correlation function of galaxies appears at
$ z_c > 1 $.  At a given $z$, the spatial correlation function for
smaller $\epsilon$ is larger than that for larger $\epsilon$.  Thus,
the angular correlation function decreases with $\epsilon$, and this
effect appears at high redshift.  In Figure 2, we show a contour of
$w_\GRB (5^{\circ})$ for model B on a $(z_c,\Delta z/z_c)$ plane.

\subsection{\smit Open and spatially-flat cosmological models}

When the background cosmological model is general or the $\xi_{gg}(x,z)$
is not simply assumed to be a power-law, one has to numerically
evaluate the integral of eq.(\ref{wxi}). The linear growth rates of
fluctuations, $D_1$, are explicitly given by
\begin{eqnarray}
D_1&=&1+\frac{3}{b_1}+\frac{3(1+b_1)^{1/2}}{{b_1}^{3/2}}
           \ln\bra{(1+b_1)^{1/2}-{b_1}^{1/2}},\\
b_1 &\equiv& (\Omega_0^{-1}-1)a/a_0 
\end{eqnarray}
for $\lambda_0=0$ models, and by
\begin{eqnarray}
\label{d1sf}
D_1&=&\sqrt{1+\frac{2}{{b_2}^3}}\int^{b_2}_0\paren{\frac{{b_2}^{\prime}}
{2+{b_2}^{\prime3}}}d{b_2}^{\prime} 
 \simeq \frac{0.2 b_2+0.044 {b_2}^{1.37}}{1+0.328 {b_2}^{1.37}}, \\
b_2 &\equiv& 2^{1/3}(\Omega_0^{-1}-1)^{1/3}a/a_0 
\end{eqnarray}
for $\lambda_0=1-\Omega_0$ models. 
For $\lambda_0=1-\Omega_0$ models, we use an empirical fitting
formula (\ref{d1sf}) in the present computation\cite{suto}.

Figures 3a and 3b illustrate $w_\GRB (5^{\circ})$ as a function of
$\Omega_0$ for models A and B, respectively (with $z_c=0.5$, and
$\Delta z/z_c=0.1$). 
When the growth rate is given independently of the value of $\Omega_0$
(as in our power-law models for $g$), the angular scale $\theta$
corresponds to the smaller physical length in the larger $\Omega_0$
model.  Thus the amplitude of $w_\GRB$ at a given $\theta$ becomes
larger for larger $\Omega_0$ since it is mainly contributed by the
spatial correlation $\xi_{gg}$ on smaller length scales.  This is not the
case when the growth rate $g$ depends on $\Omega_0$.  Since we
normalize the amplitude of $\xi_{gg}$ at the present epoch, lower
$\Omega_0$ models, especially $\lambda_0=0$ cases, should have higher
amplitude of $\xi_{gg}$ in the past (since linear density fluctuations do
not grow for $z \simlt \Omega_0^{-1}$.)  This is why the amplitude of
$w_\GRB$ decreases with increasing $\Omega_0$ (for $\lambda_0=0$
cases) when we adopt linear growth rate for $
g(z,\Omega_0,\lambda_0)$.  In any case, the variation of the predicted
amplitude of $w_\GRB$ for different is very small, and at most within
a factor of 2.

\subsection{\smit CDM spectrum}

So far we adopted power-law models (eq.[\ref{poxi}]) for the spatial
correlations of galaxies. As an example based on fairly
specific theoretical models,
let us
consider the cold dark matter (CDM) models assuming that galaxies
trace mass (dark matter). In this case, we adopt $\xi(x,z)_{gg}$
computed from the CDM power spectrum $P(k,z)$ in linear theory:
\begin{equation}
\xi_{gg} (x,z)= {1 \over 2\pi^2 }
                  \int P(k,z) { \sin kx \over kx} k^2 dk .
\label{xipk}
\end{equation}
As for $P(k,z)$, we use a fitting formula for the primordial
Harrison-Zel'dovich spectrum\cite{bbks}:
\begin{equation}
P(k,z) = C_0 D_1(z) k 
   \left[{ {\rm ln}~(1+2.34q) \over 2.34q} \right]^2
\left[1+3.89q + (16.1q)^2 + (5.46q)^3 + (6.71q)^4 \right]^{-1/2} ,
                           \label{cdmpk}
\end{equation}
where $q \equiv k/ (\Omega_0h^2 {\rm Mpc}^{-1})$, and the amplitude
$C_0$ is fixed so that the top-hat filtered mass fluctuation becomes
unity at $r=8\himpc$ (see, e.g. Ref.8).

We evaluate $\xi_{gg}(x,z=0)$ and $w_{\GRB}(\theta)$ numerically in the
CDM model, which are shown in Figures 4 and 5 for model A in the
Einstein-de Sitter model ($\Omega_0=1.0$, $\lambda_0=0.0$ and
$h=0.5$).  The amplitude of $w_\GRB(\theta)$ for the CDM model rapidly
decreases if the angle is larger than the angle spanning the horizon
scale at the equality epoch which locates at $z \simeq z_c$.  This is
because the amplitude of $\xi_{gg}(x,z=0)$ for the CDM model decreases
rapidly outside the horizon scale at the equality epoch.  Incidentally
this may provide useful information of $z_c$ and $\Omega_0$, if the
angular scale of the zero-point of $w_\GRB(\theta)$ is
detectable\cite{kr}.

\section{\smbf Discussion and conclusions}

It is important to ask here whether or not the amplitude of
$w_\GRB(\theta)$ predicted in the previous section is detectable.
This is checked as follows.  The expected number of GRB pairs at
separation $\theta$ to $\theta + \delta \theta$ is
\begin{equation}
\langle DD(\theta)\rangle \delta \theta
 = \frac{N_\GRB}{2} \frac{N_\GRB}{4 \pi} [1+w_\GRB(\theta)]
  2 \pi \theta \delta \theta ,
\label{ddtheta}
\end{equation}
where $N_\GRB$ is the number of observed GRBs homogeneously surveyed
over the $4\pi$ steradian.  A statistically significant estimate of
$w_\GRB(\theta)$ requires that the number of pairs in excess of random
(the second term in RHS of eq.[\ref{ddtheta}]) should be at least larger than
the shot noise fluctuation given by
\begin{equation}
\langle \Delta DD(\theta)\rangle \delta \theta
 = \sqrt{\frac{N_\GRB}{2} \frac{N_\GRB}{4 \pi} 
         \, 2 \pi \theta \delta \theta} .
\end{equation}
This implies that given $w_\GRB(\theta)$, $\theta$, and $\delta
\theta$, one needs
\begin{equation}
N_\GRB \geq 2300 \left({10^{-2} \over w_\GRB(\theta)}\right)
\sqrt{5^\circ \over \theta} \sqrt{5^\circ \over \delta\theta} .
\end{equation}
The lower bound on $N_\GRB$ for successful detection of
$w_\GRB(5^\circ)$ is listed in the last row in Table 1.  Taking into
account the BATSE detection rate of GRBs $\sim 1$ per day\cite{kov},
the amplitude of $w_\GRB$ in most models is rather small for
successful detection. Nevertheless it is marginally detectable in some
models (for instance, model A with $ z_c \leq 0.3 $,
model B with $ z_c=1.0, \Delta z/z_c = 0.1 $ and $ \epsilon=-3+\gamma$)
with GRBs observed by BATSE in several years.

This difficulty in detecting $w_\GRB$ is partly ascribed to the low
angular resolution of the currently available $\gamma$-ray detectors.
In fact, it would be much easier to detect $w_\GRB$ at small angular
separations for a given number of sources; $ w_\GRB(\theta) \propto
\theta^{1-\gamma} $ for power-law model (see Fig. 4).
The correlation of GRBs at $1^{\circ} $ is about three times as large
as that at $ 5^{\circ} $, and the necessary number of GRBs reduces
factor of 1.4. In this respect, radio, optical and X-ray follow-up
observations, if successful, would provide us very valuable
information on $w_\GRB$.

Recently, Cohen, Kolatte and Piran\cite{coh} reported a possible
association of GRBs with rich clusters.  As long as neutron stars are
involved in GRB events, this association would be difficult to
explain. Therefore their report is a serious challenge for the GRB
mechanism with neutron stars. On the other hand, if their claim turns
out to be real, the amplitude of $w_\GRB$ would become significantly
larger than computed in the previous section since clusters of
galaxies are known to be much more strongly clustered than galaxies;
depending on the richness of clusters which GRBs are associated with,
the amplitude would be enhanced by even an order of magnitude.  Then
$w_\GRB(5^\circ)$ would be detectable in a few years for most
evolutionary models considered here.


In summary, we employ several simple models for the evolution of
galaxy correlation functions, and compute the angular two-point
correlation functions of the GRBs assuming that the GRB event rate is
constant. On the basis of the plausible evolutional models for galaxy
clustering up to $z \simlt 1$, the predicted amplitudes of $w_\GRB$ at
$5^\circ$ scale are generally too small to be detected with the
current $\gamma$-ray instruments, although some (optimistic) models
predict marginally detectable values.  Such small amplitudes are
mainly due to our simplifying assumption that the GRBs faithfully
reflect the clustering at high redshift galaxies. Without any other
convincing theory, we believe that this assumption is most plausible
and realistic. Nevertheless the amplitudes of the angular correlation
functions may become significantly larger than our predictions either
if the GRBs are associated with highly biased species of galaxies
(including a possible association of rich clusters) or if the galaxy
clustering around $z\simlt 1$ is much stronger than what standard
cosmological models predict. Obviously either case seems rather
unrealistic, but cannot be ruled out until specific GRB model is
established.

Therefore even if no positive signature of $w_\GRB$ around $5^\circ$
scales is detected from future data in several years, this is
completely consistent with their cosmological models. In turn, in case
that $w_\GRB$ are detected, the result provides us useful information
on the correlation of galaxies at high redshifts provided that the
GRBs are cosmological.

\vspace{1cm}

One of authors (S.K.)  acknowledges the support by the Fellowships of
Japan Society for the Promotion of Science for Japanese Junior
Scientists.  This research was supported in part by the Grants-in-Aid
by the Ministry of Education, Science and Culture of Japan (05640312,
06233209).

\bigskip

\parskip=0pt
\bibliographystyle{plain}

\newpage

\begin{center}

Table 1. Summary of $w_\GRB(5^{\circ})$.

\medskip
\begin{tabular}{ccccccrr} \hline
model & $z_c$ & $\Delta z/z_c$ & $\Omega_0$ & $\lambda_0$ &$\epsilon$
    & $w_\GRB(5^\circ)$ & $N_\GRB$ \\\hline
A& 0.5 & -- & 1.0 & 0.0 & 0 &  $ 2.9\times 10^{-3}$ & 7900 \\
A& 1.0 & -- & 1.0 & 0.0 & 0 &  $ 9.7\times 10^{-4}$ & 24000 \\
B& 0.5 & 0.1& 1.0 & 0.0 & 0 &  $ 1.7\times 10^{-2}$ & 1400 \\
B& 1.0 & 0.1& 1.0 & 0.0 & 0 &  $ 6.2\times 10^{-3}$ & 3700 \\
A& 0.5 & -- & 1.0 & 0.0 & -3+$\gamma$ &  $ 4.3\times 10^{-3}$ & 5300 \\
A& 1.0 & -- & 1.0 & 0.0 & -3+$\gamma$ &  $ 1.9\times 10^{-3}$ & 12000 \\
B& 0.5 & 0.1& 1.0 & 0.0 & -3+$\gamma$ &  $ 2.7\times 10^{-2}$ & 800 \\
B& 1.0 & 0.1& 1.0 & 0.0 & -3+$\gamma$ &  $ 1.5\times 10^{-2}$ & 1500 \\\hline
A& 0.5 & -- & 0.2 & 0.0 & 0 &  $ 2.5\times 10^{-3}$ & 9200 \\
A& 1.0 & -- & 0.2 & 0.0 & 0 &  $ 7.5\times 10^{-4}$ & 31000 \\
B& 0.5 & 0.1& 0.2 & 0.0 & 0 &  $ 1.3\times 10^{-2}$ & 1700 \\
B& 1.0 & 0.1& 0.2 & 0.0 & 0 &  $ 4.2\times 10^{-3}$ & 5500 \\
A& 0.5 & -- & 0.2 & 0.0 & -3+$\gamma$ &  $ 3.7\times 10^{-3}$ & 6200 \\
A& 1.0 & -- & 0.2 & 0.0 & -3+$\gamma$ &  $ 1.5\times 10^{-3}$ & 15000 \\
B& 0.5 & 0.1& 0.2 & 0.0 & -3+$\gamma$ &  $ 2.2\times 10^{-2}$ & 1100 \\
B& 1.0 & 0.1& 0.2 & 0.0 & -3+$\gamma$ &  $ 9.8\times 10^{-3}$ & 2300 \\\hline
A& 0.5 & -- & 0.2 & 0.8 & 0 &  $ 1.9\times 10^{-3}$ & 12000 \\
A& 1.0 & -- & 0.2 & 0.8 & 0 &  $ 5.1\times 10^{-4}$ & 45000 \\
B& 0.5 & 0.1& 0.2 & 0.8 & 0 &  $ 9.3\times 10^{-3}$ & 2500 \\
B& 1.0 & 0.1& 0.2 & 0.8 & 0 &  $ 2.6\times 10^{-3}$ & 8800 \\
A& 0.5 & -- & 0.2 & 0.8 & -3+$\gamma$ &  $ 2.9\times 10^{-3}$ & 7900 \\
A& 1.0 & -- & 0.2 & 0.8 & -3+$\gamma$ &  $ 1.0\times 10^{-3}$ & 23000 \\
B& 0.5 & 0.1& 0.2 & 0.8 & -3+$\gamma$ &  $ 1.5\times 10^{-2}$ & 1500 \\
B& 1.0 & 0.1& 0.2 & 0.8 & -3+$\gamma$ &  $ 6.2\times 10^{-3}$ & 3700 \\\hline
\end{tabular}
\end{center}

\bigskip
\bigskip

\newpage
\noindent {\bf Figure captions}
\bigskip

\pp {\bf Figure 1.}
The angular two-point correlation functions of GRBs at $\theta=5^{\circ}$ as a
function of $z_c$ in the Einstein -- de Sitter model.  The
thick solid, dotted and thin solid lines represent results for $
\epsilon=0$, $\epsilon = -3 + \gamma $ and linear growth rate,
respectively.  (a) model A; (b) model B with $ \Delta z/z_c =0.1$.

\pp {\bf Figure 2.}
The contourlines of $w_{GRB}(5^{\circ})$ in the ($z_c, \Delta z/z_c$) plane
for model B (the Einstein -- de Sitter model).  The thick solid,
dotted and thin solid lines represent results for $ \epsilon=0$, $
\epsilon = -3 + \gamma $ and linear growth rate, respectively.

\pp {\bf Figure 3.}
The angular correlation $w_{GRB}(5^{\circ})$ plotted against $ \Omega_0 $.
Thick lines indicate results for models without cosmological constant,
while thin lines for spatially flat models ($\lambda_0=1-\Omega_0$).
Solid, dotted, and dashed lines correspond to $ \epsilon=0$
and $-3 + \gamma$, and linear growth rate, respectively.  (a) model A
with $z_c=0.5$, (b) model B with $ z_c=0.5 $ and $ \Delta z/z_c =
0.1$.

\pp {\bf Figure 4.}
The two-point correlation function of galaxies plotted against separation
$x$ (the Einstein-de Sitter model). Solid line corresponds to the CDM
model ($\Omega_0=1.0, \lambda_0=0.0 $ and $h=0.5$),
while dotted line corresponds to a power-law model.

\pp {\bf Figure 5.}
The angular two-point correlation function of GRBs plotted against
angular separation $ \theta $.  Thick and thin lines indicate results
for CDM model and for power-law model. In both cases, we adopt model A
for $\phi(z)$ with $z_c=1.0$ and $0.1$, and linear growth rate in the
Einstein-de Sitter model for $g$.

\newpage
\epsfile{file=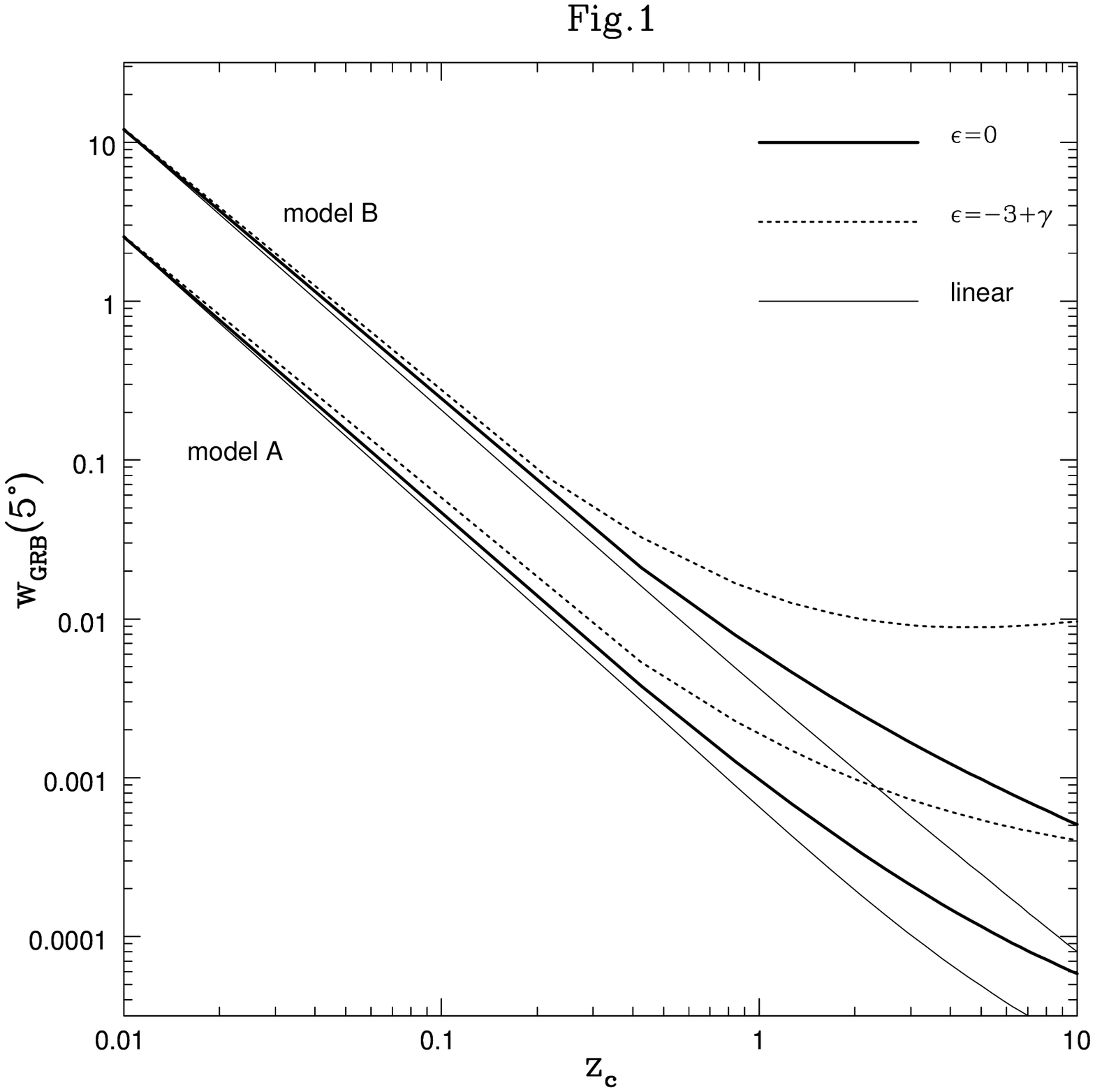,width=16cm}

{\bf
\noindent
Kobayashi, Sasaki and Suto \\
Angular two-point correlation functions \\
for cosmological gamma-ray burst model \\
Fig.1}

\newpage
\epsfile{file=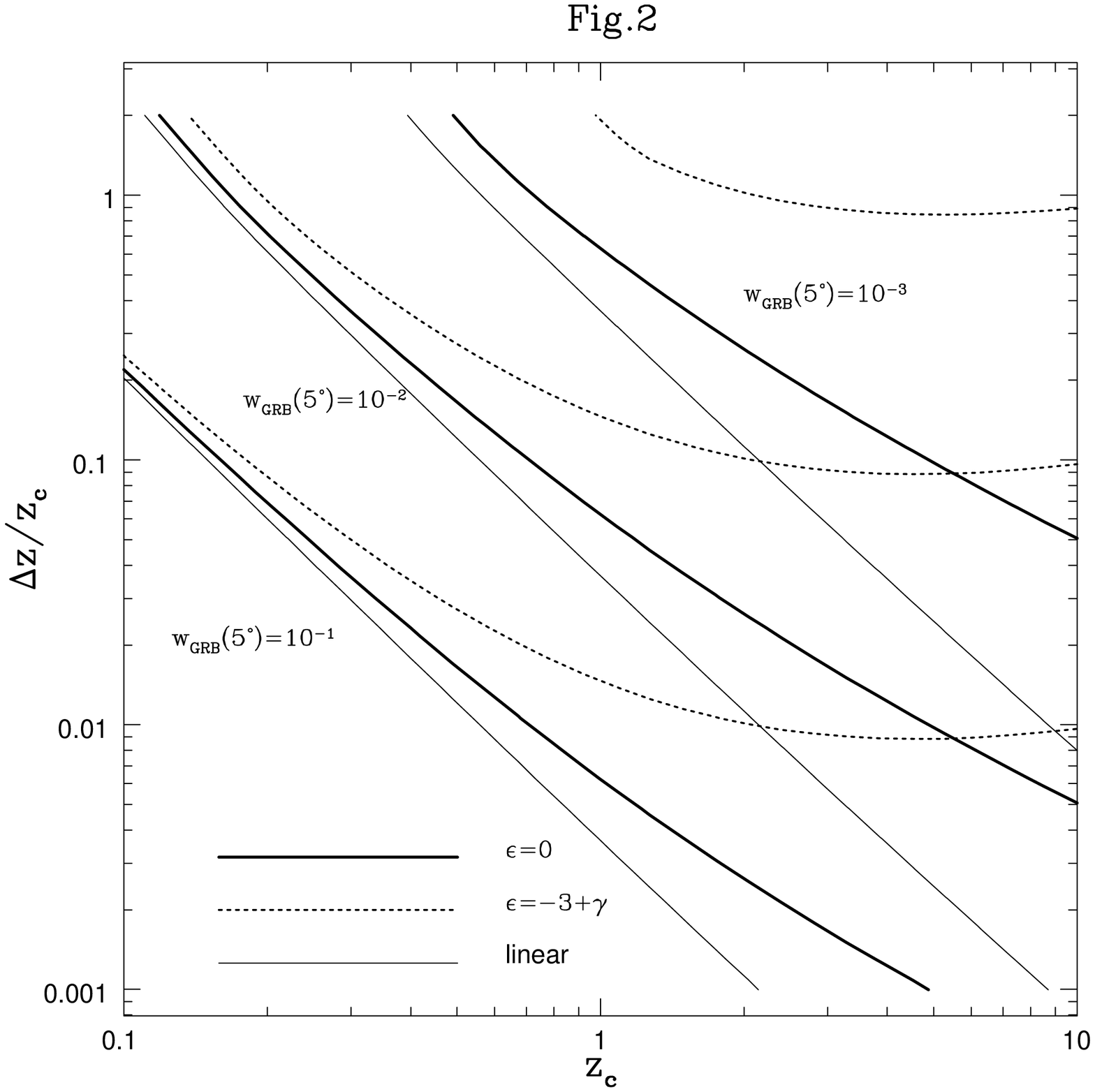,width=16cm}

{\bf
\noindent
Kobayashi, Sasaki and Suto \\
Angular two-point correlation functions \\
for cosmological gamma-ray burst model \\
Fig.2}
\newpage
\epsfile{file=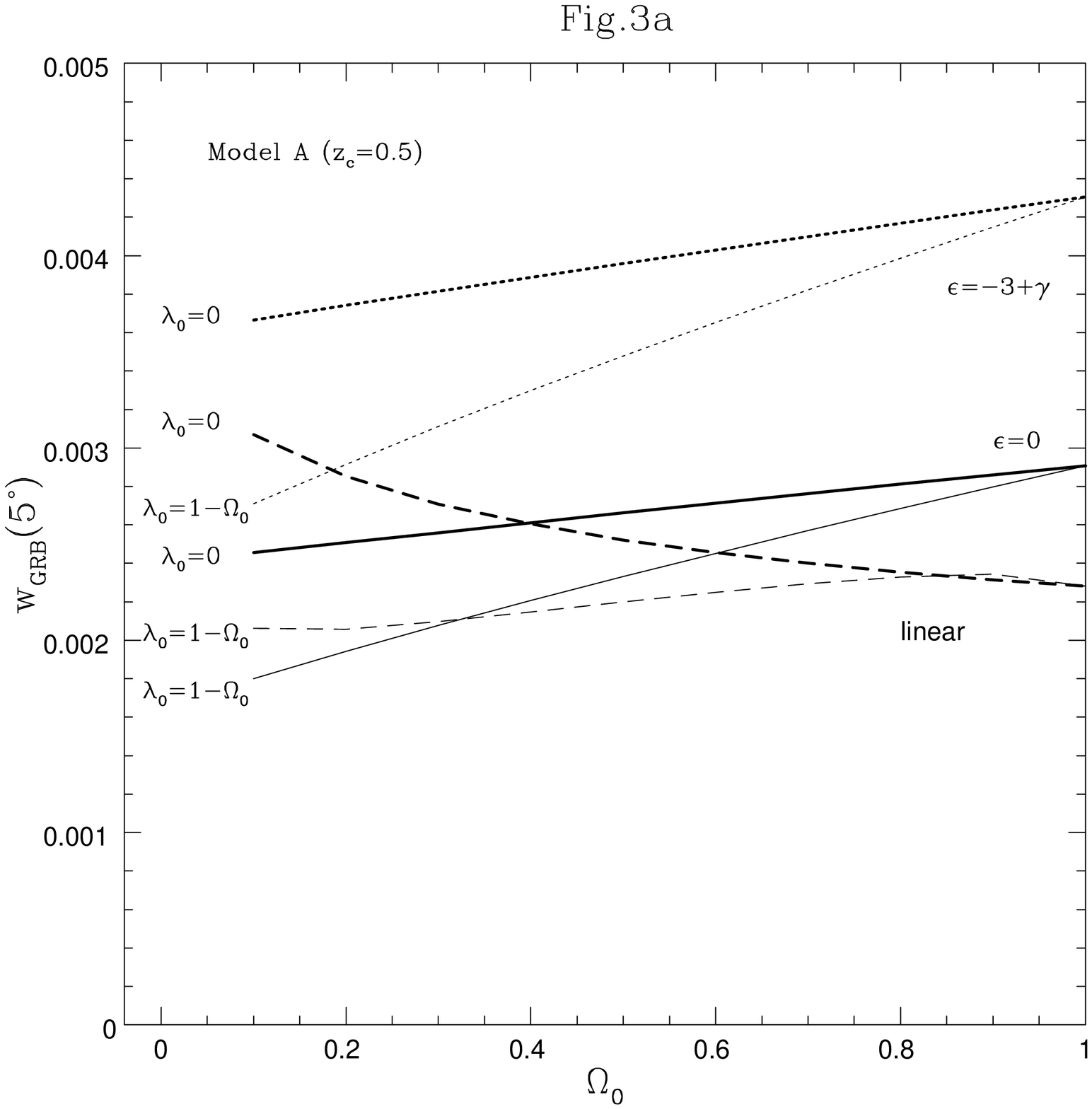,width=16cm}

{\bf
\noindent
Kobayashi, Sasaki and Suto \\
Angular two-point correlation functions \\
for cosmological gamma-ray burst model \\
Fig.3a}
\newpage
\epsfile{file=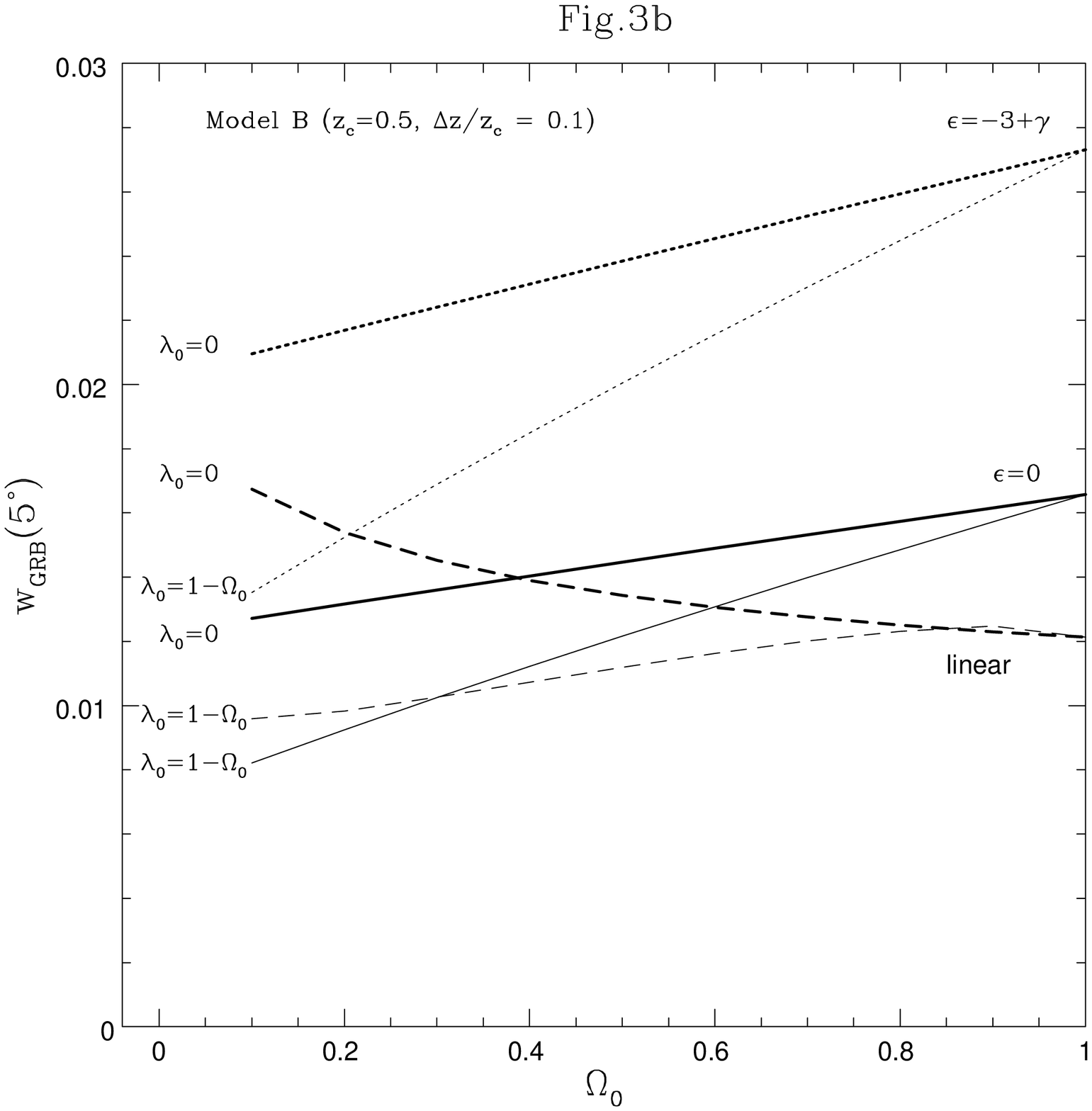,width=16cm}

{\bf
\noindent
Kobayashi, Sasaki and Suto \\
Angular two-point correlation functions \\
for cosmological gamma-ray burst model \\
Fig.3b}
\newpage
\epsfile{file=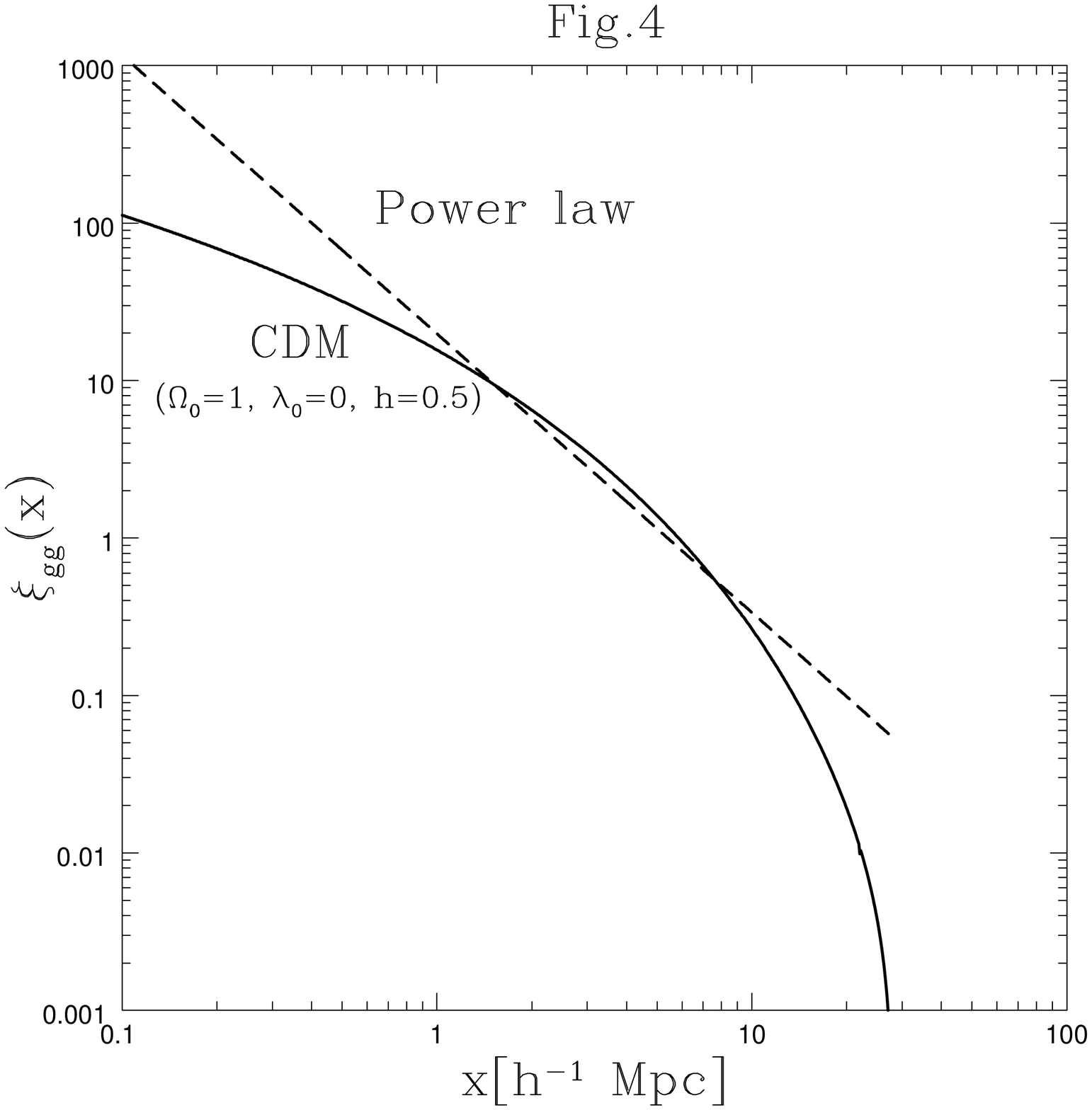,width=16cm}

{\bf
\noindent
Kobayashi, Sasaki and Suto \\
Angular two-point correlation functions \\
for cosmological gamma-ray burst model \\
Fig.4}
\newpage
\epsfile{file=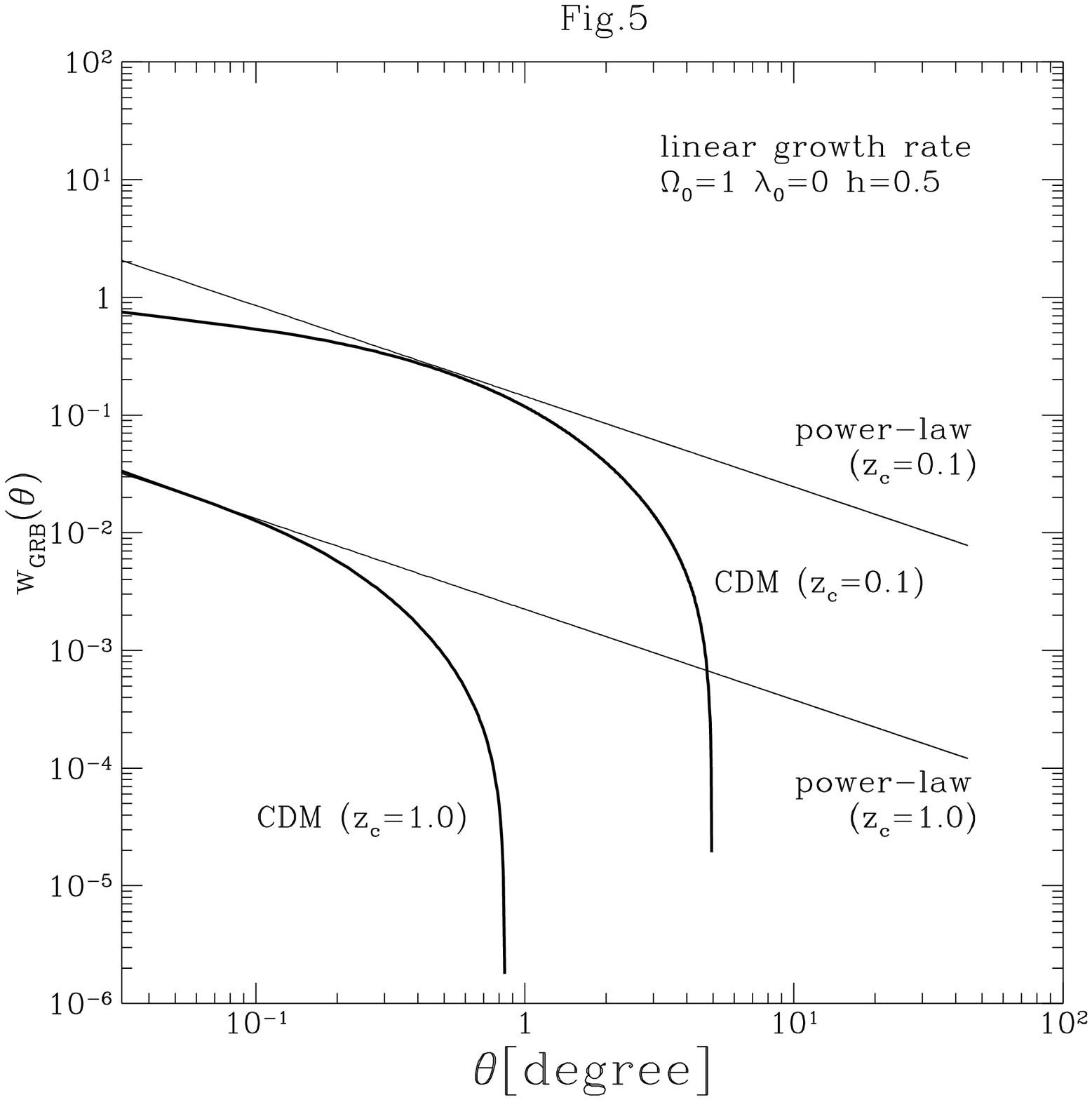,width=16cm}

{\bf
\noindent
Kobayashi, Sasaki and Suto \\
Angular two-point correlation functions \\
for cosmological gamma-ray burst model \\
Fig.5}

\end{document}